\documentclass[aps,pra,twocolumn,superscriptaddress]{revtex4}
\usepackage{amsfonts}
\usepackage{amssymb}
\usepackage{amsmath}
\usepackage{epsfig}
\usepackage{color}
\usepackage{graphics, graphicx}
\usepackage{bbold}
\usepackage{psfrag}
\usepackage{mathcomp}
\usepackage{subfigure}
\usepackage{verbatim}
\usepackage{color}
\usepackage[colorlinks,citecolor=blue]{hyperref}

\begin{document}

\title{Synthetic Hall tube of interacting fermions}
\author{Xiaofan Zhou}
\affiliation{State Key Laboratory of Quantum Optics and Quantum Optics Devices, Institute
of Laser spectroscopy, Shanxi University, Taiyuan 030006, China}
\author{Gang Chen}
\email{chengang971@163.com}
\affiliation{State Key Laboratory of Quantum Optics and Quantum Optics Devices, Institute
of Laser spectroscopy, Shanxi University, Taiyuan 030006, China}
\affiliation{Collaborative Innovation Center of Extreme Optics, Shanxi University,
Taiyuan, Shanxi 030006, China}
\affiliation{Collaborative Innovation Center of Light Manipulations and Applications,
Shandong Normal University, Jinan 250358, China}
\author{Suotang Jia}
\affiliation{State Key Laboratory of Quantum Optics and Quantum Optics Devices, Institute
of Laser spectroscopy, Shanxi University, Taiyuan 030006, China}
\affiliation{Collaborative Innovation Center of Extreme Optics, Shanxi University,
Taiyuan, Shanxi 030006, China}

\begin{abstract}
Motivated by a recent experiment [J. H. Han, \textit{et. al.}, Phys. Rev.
Lett.~\textbf{122}, 065303 (2019)], we investigate many-body physics of
interacting fermions in a synthetic Hall tube, using state-of-the-art density-matrix renormalization-group numerical method.
Since the inter-leg couplings of this synthetic Hall tube generate an interesting spin-tensor Zeeman field, exotic topological and magnetic properties occur. Especially, four new quantum phases, such as nontopological spin-vector and -tensor paramagnetic insulators, and topological and nontopological spin-mixed paramagnetic insulators, are predicted by calculating entanglement spectrum, entanglement
entropies, energy gaps, and local magnetic orders with 3 spin-vectors and 5 spin-tensors. Moreover, the topologically magnetic phase transitions induced by the interaction
as well as the inter-leg couplings are also revealed. Our results pave a new way to explore many-body (topological) states induced by both the spiral spin-vector and -tensor Zeeman fields.
\end{abstract}

\maketitle

\section{Introduction}

Since the discovery of the quantum Hall effect~\cite{Hall}, the exploration of novel topological states of matter has been attracted great attention in both theory and experiment,
since they provide important applications in designing novel quantum devices and processing quantum information.
The Hofstadter-Harper Hamiltonian is one of the fundamental models that are used to investigate topological states~\cite%
{Hofstadter1976}. The experimental realization of such Hamiltonian in cold atomic gases opens up the avenue of simulating
topological states~\cite{Goldman2016,Tai2017,Cooper2019}. In cold
atoms system, the internal degrees of freedom of atoms, such as the
hyperfine spins~\cite{Mancini2015,Stuhl2015} and clock states~\cite%
{Livi2016,zhou2017current}, can be treated as synthetic dimension to simulate
the $D$+1 dimensional quantum physics using $D$ dimensional lattices~\cite%
{Boada2012}, e.g., four dimensional quantum Hall effect~\cite{Price2015} and
chiral edge current of Hall ribbons~\cite{Mancini2015,Stuhl2015}.

Using three hyperfine states as synthetic lattice dimension and coupling
them through synthetic gauge fields by two-photon Raman process~\cite%
{SDprl14}, spin-1 spin-orbit coupling~\cite%
{Natu2015,Barbarino2015,spin-1SOC2016,SOC-1BEC1,Pixley2017} and spin-tensor-momentum
coupling~\cite{Luo2017,Li2020} have also been implemented. When the links between the hyperfine states are cyclic with a gauge flux $\phi
=2\pi /3$, the optical lattice can form a synthetic Hall tube~\cite%
{Barbarino2018,Luo2020}, which is a simple Hofstadter-Harper Hamiltonian~%
\cite{Hofstadter1976}. The synthetic Hall tube supports a generalized
inversion symmetry-protected topological insulator~\cite%
{Nourse2016}, which is similar as the integer quantum Hall
state. Since the time-reversal, particle-hole,
and chiral symmetries are broken, this topological insulator belongs to the symmetry class A (unitary) of the Altland-Zirnbauer
classification~\cite{Altland1997,Schnyder2008,Ludwig2016,Chiu2016}. As varying one of the inter-leg coupling strength, there
exists a topological phase transition with a closing band gap at critical point
~\cite{Barbarino2018}. In a recent experiment, this interesting synthetic Hall tube has been realized successfully in the alkaline-earth fermions~\cite{Han2019}.

Apart from the single-particle quantum
regulation in cold atoms experiments, the interactions between the internal states can be controlled
via Feshbach resonances~\cite{Chin2010}, and more importantly,
generate rich many-body phenomena~\cite{Raghu2008,Bloch2008,Mueller2017,Rachel2018,Junemann2017,zhou2017SPT,Leseleuc2019}. However, the interacting synthetic Hall tube has not been fully investigated. In this paper, we investigate many-body properties of such system, based on state-of-the-art density-matrix renormalization-group (DMRG) numerical method~\cite{dmrg1,dmrg2}. Since the inter-leg couplings of this synthetic Hall tube generate an interesting spin-tensor Zeeman field, it is very necessary to explore magnetic properties of the system, apart from the interaction-driven topological transition. Due to the coexistence of the spiral spin-vector and -tensor Zeeman fields in the synthetic Hall tube, local magnetic orders with 3 spin-vectors and 5 spin-tensors should be introduced \cite{zhou2020}. In terms of the calculated entanglement spectrum, entanglement
entropies, energy gaps, and local magnetic orders, we find four new quantum phases such as nontopological spin-vector and -tensor paramagnetic insulators, and topological and nontopological spin-mixed paramagnetic insulators. Moreover, the topologically magnetic phase transitions induced by the interaction
as well as the inter-leg couplings are also revealed. Our results pave a new way to explore many-body (topological) states induced by both the spiral spin-vector and -tensor Zeeman fields.\newline

\begin{figure}[t]
\centering
\includegraphics[width = 3.2in]{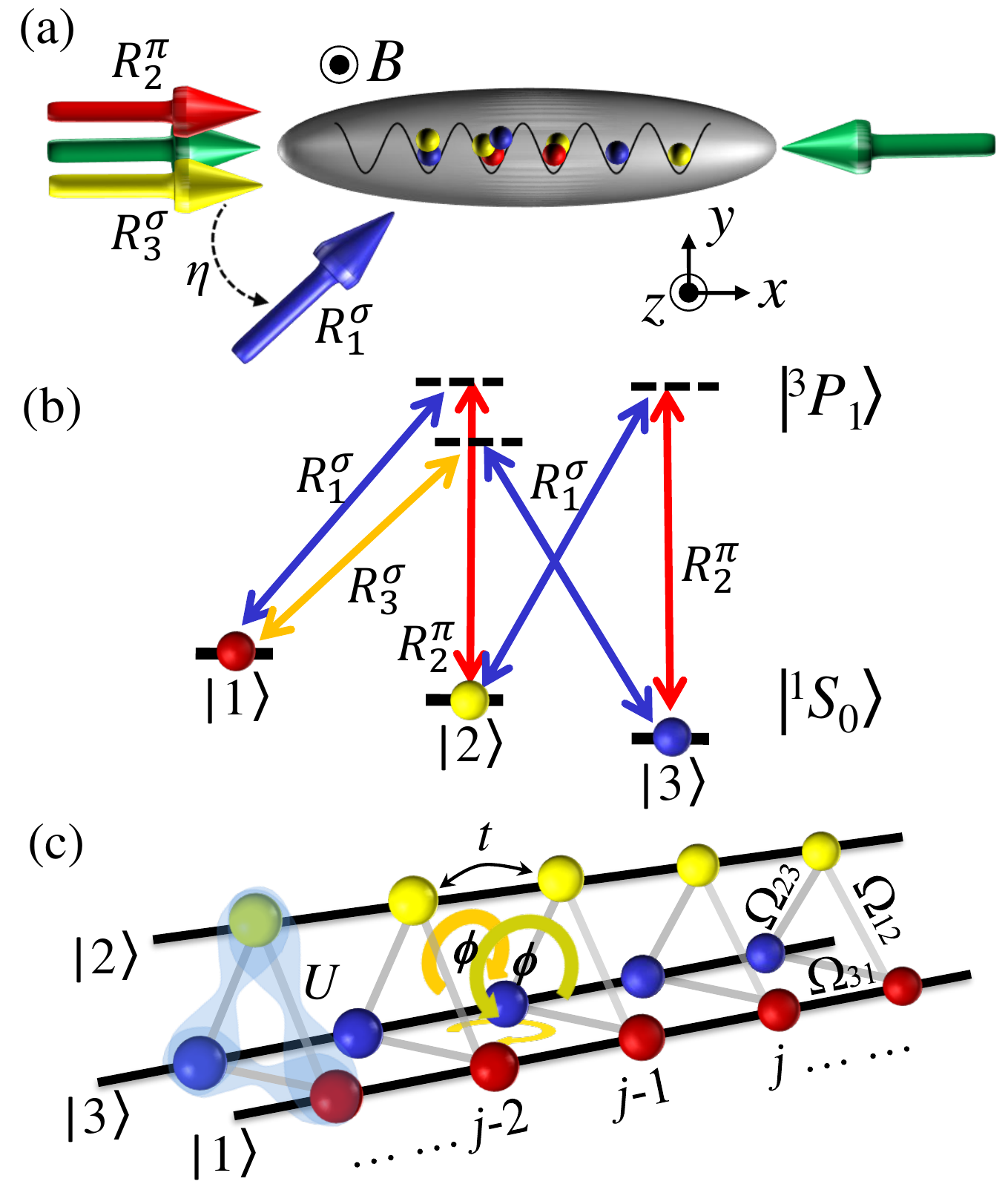} \hskip 0.0cm
\caption{(a) Schematics of the system setup with three Raman lasers $%
R_{1,2,3}^{\protect\sigma ,\protect\pi ,\protect\sigma }$, which are
represented respectively by the yellow, blue, and red arrows. $R_{1}^{%
\protect\sigma }$ has an angle $\protect\eta $ from $x$-axis. A magnetic
field $B$ along $z$-axis is applied to lift the spin degeneracy of the
ground state $|^{1}S_{0}\rangle $. (b) Three hyperfine spin states in ground
state $|^{1}S_{0}\rangle $ of alkaline-earth(-like) atoms $^{173}$Yb are
coupled by three two-photon Raman transitions. (c) Synthetic Hall tube with
a uniform flux $\protect\phi $ on each side plaquette and interaction $U$
between these hyperfine spin states.}
\label{fig:exper}
\end{figure}

\section{Model and Hamiltonian}

\label{Model and Hamiltonian}

Similar as Ref.~\cite{Han2019}, here we consider the alkaline-earth fermions
$^{173}$Yb trapped in an effective one-dimensional optical lattice (along $x$%
-direction), as shown in Fig.~\ref{fig:exper}(a). Three hyperfine spin
states of the ground state $|^{1}S_{0}\rangle $, $|1\rangle
=|F=5/2,m_{F}=-5/2\rangle $, $|2\rangle =|F=5/2,m_{F}=-3/2\rangle $ and $%
|3\rangle =|F=5/2,m_{F}=-1/2\rangle $, are chosen as three legs, as shown in
Fig.~\ref{fig:exper}(b). Three linearly-polarized Raman laser beams $%
R_{1,2,3}^{\sigma ,\pi ,\sigma }$ are used to make three two-photon Raman
transitions between the states $|^{1}S_{0},F=5/2\rangle $ and $%
|^{3}P_{1},F=7/2\rangle $. The couplings $|1\rangle \leftrightarrow
|2\rangle $ and $|2\rangle \leftrightarrow |3\rangle $ are the $\pi -\sigma $
transitions ($\Delta m_{F}=1$), while the coupling $|1\rangle
\leftrightarrow |3\rangle $ is the $\sigma -\sigma $ transition ($\Delta
m_{F}=2$), as shown in Fig.~\ref{fig:exper}(b). Thus, the three-component
atomic tunneling along the lattice and three-leg couplings with a complex
phase factor form a synthetic tube with a uniform flux per plaquette, as
shown in Fig.~\ref{fig:exper}(c).

When the effective one-dimensional optical
lattice is deep enough and the Rabi frequency of the two-photon Raman
transitions is not too large, we use the single-band approximation to derive
the tight-binding model Hamiltonian~\cite{Han2019}
\begin{equation}
\hat{H}=\hat{H}_{\text{hop}}+\hat{H}_{\Omega }+\hat{H}_{\text{int}},
\label{eq:tb}
\end{equation}%
where the tunneling Hamiltonian
\begin{equation}
\hat{H}_{\text{hop}}=\sum_{j,\sigma }(-t\hat{c}_{j+1,\sigma }^{\dagger }\hat{%
c}_{j,\sigma }+\text{H.c.}),  \label{Ht}
\end{equation}%
the inter-leg coupling Hamiltonian
\begin{equation}
\hat{H}_{\Omega }=\frac{1}{2}\sum_{j,\sigma \neq \sigma ^{\prime }}(\Omega
_{\sigma \sigma ^{\prime }}e^{i\phi j}\hat{c}_{j,\sigma }^{\dagger }\hat{c}%
_{j,\sigma ^{\prime }}+\text{H.c.}),  \label{H_Omega}
\end{equation}%
and the interaction Hamiltonian
\begin{equation}
\hat{H}_{\text{int}}=U\!\!\!\sum_{j,\sigma \neq \sigma ^{\prime }}\!\!\hat{n}%
_{j,\sigma }\hat{n}_{j,\sigma ^{\prime }}.  \label{Hint}
\end{equation}%
In Eq.~(\ref{eq:tb}), $\hat{c}_{j,\sigma }$ ($\hat{c}_{j,\sigma }^{\dagger }$%
) is the annihilation (creation) operator for a fermion at the real lattice
site $j=1,\cdots ,L$ with spin $\sigma =\left( 1,2,3\right) $ and the
lattice length $L$, $\hat{n}_{j,\sigma }\equiv \hat{c}_{j,\sigma }^{\dagger }%
\hat{c}_{j,\sigma }$ is the number operator. $t$ is the tunneling rate, $%
\Omega _{\sigma \sigma ^{\prime }}$ is the Rabi frequency of the two-photon
Raman transition between the spin states $\left\vert \sigma \right\rangle $
and $\left\vert \sigma ^{\prime }\right\rangle $ and is set to $\Omega
_{12}=\Omega _{23}$ for simplicity, the $j$-dependent complex phase factor $%
e^{i\phi j}$ results from the momentum imparted by the two-photon Raman
transitions, the flux $\phi =k_{R}d_{x}(1-\cos \eta )$ with $k_{R}$ being
the recoil momentum of the Raman lasers and $d_{x}$ being the lattice
constant, $U$ is the interaction strength, and H.c.~is the Hermitian
conjugate.

The Hamiltonian (\ref{eq:tb}) has a distinct advantage that all
parameters can be tuned independently. For example, $t$ can be tuned by
varying the depth of the optical lattice, $\Omega _{\sigma \sigma ^{\prime }}
$ can be controlled by adjusting the magnitudes of the Raman laser beams, $%
\phi $ can be manipulated by controlling the angle $\eta $, and $U$ can be
tuned via the external magnetic field through orbital Feshbach resonance~%
\cite{ren1,ofrexp1,ofrexp2} or via the transverse trapping frequencies
through the confinement induced resonance~\cite{Zhang2016,Bergeman2003}. In
the following, we mainly consider the case of the half filling, i.e., $%
n=N/L=1$ with $N$ being the total number of atoms, since the system exhibits
a synthetic Hall tube in such condition. We also address the repulsive
interaction $U>0$ and set $t=1$ as a unit.

In the absence of interaction ($U=0$), when $\phi =2\pi /3$ and $\Omega
_{-}<\Omega _{31}<\Omega _{+}$ with $\Omega _{\pm }=\pm 3t+\sqrt{\Omega
_{12}^{2}+9t^{2}}$, this synthetic Hall tube supports a topological
insulator protected by generalized inversion symmetry \cite%
{Nourse2016,Barbarino2018,Han2019}. Since the time-reversal, particle-hole,
and chiral symmetries are broken, the topological insulator belongs to the
unitary symmetry class A (unitary) of the Altland-Zirnbauer classification
and is characterized by a $\mathbb{Z}$ invariant~\cite%
{Altland1997,Schnyder2008,Ludwig2016,Chiu2016}. More interestingly, the
inter-leg couplings generate spatially periodic spin-vector and -tensor
Zeeman fields with the following Hamiltonian%
\begin{equation}
\begin{split}
\hat{H}_{\Omega }& =\sum_{j}\Omega _{12}\left[ \cos (\phi j)S_{j}^{x}-\sin
(\phi j)S_{j}^{y}\right] \\
& +\Omega _{31}\left[ \cos (\phi j)\left( N_{j}^{xx}-N_{j}^{yy}\right) +\sin
(\phi j)N_{j}^{xy}\right] ,
\end{split}
\label{HOmega}
\end{equation}%
where $\mathbf{S}_{j}=\sum_{\sigma \sigma ^{\prime }}{b}_{j\sigma }^{\dag }%
\mathbf{F}_{\sigma \sigma ^{\prime }}{b}_{j\sigma ^{\prime }}$ with $\mathbf{%
F}_{\sigma \sigma ^{\prime }}$ being the spin operators of the total angular
momentum $F=1$, $N^{\alpha \beta }=\{S^{\alpha },S^{\beta }\}/2-\delta
_{\alpha \beta }\mathbf{S}^{2}/3$ with being the anticommutation
relation and $\alpha (\beta )=(x,y,z)$, $\Omega_{12}$ and $\Omega_{31}$ are called
the spin-vector and -tensor Zeeman fields respectively, and $2\pi /\phi $ is the
spiral period of the Zeeman field. When $\Omega_{31}=0$, the synthetic Hall tube reduces to the spin-1 spin-orbit coupled optical lattice
only with the spin-vector Zeeman field \cite%
{Natu2015,Barbarino2015,spin-1SOC2016,SOC-1BEC1}, which has a trivial
topology. Notice that $\Omega_{12}$ can also be treated as the spin-tensor Zeeman field since the Hamiltonian (\ref{HOmega}) has the rotational symmetry.

In the presence of weak interaction, the topological insulator with the $%
\mathbb{Z}$ invariant still exists since the generalized inversion symmetry
remains~\cite{Morimoto2015}. When further increasing the interaction
strength, the topology of the system becomes trivial. On the other hand, the interaction Hamiltonian (\ref{Hint}) can also be rewritten as a
magnetic form
\begin{equation}
\hat{H}_{\text{int}}=\frac{U}{2}\sum_{j}{S_{j}^{z}}^{2}(\hat{n}_{j}-1).
\label{HU}
\end{equation}%
It shows clearly that at the half filling $\left( n=1\right) $, the
interaction has a little contribution on the magnetism of the system, i.e.,
the magnetic properties of the synthetic Hall tube are mainly determined by $%
\hat{H}_{\Omega }$.

Based on above qualitative analysis, it can be found that the synthetic Hall
tube exhibits exotic topological and magnetic properties arising from the
competition between the tunneling, spin-vector and -tensor Zeeman fields,
and interaction. In order to quantitatively reveal them, we will perform
state-of-the-art DMRG numerical method, for which we retain $400$ truncated
states per DMRG block and perform $30$ sweeps with a maximum truncation
error $\sim 10^{-10}$.\newline

\begin{figure*}[t]
\centering
\includegraphics[width = 7.0in]{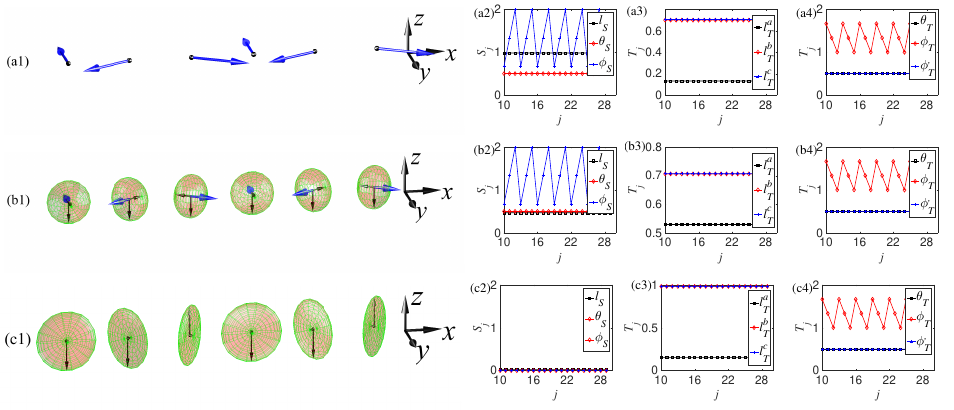} \hskip 0.0cm
\caption{(a1, b1, c1) Schematic diagrams of the spin-vector density arrows $%
\vec{S_{j}}$ and the spin-tensor density ellipsoids $T_{j}$. The blue arrow
denotes the spin-vector $\vec{S}$, while the red ellipsoid reflects the
spin-tensor $T$, in which the black arrows are the ellipsoid's axis
orientations. (a2, b2, c2) Spatial distributions of $\left[ l_{S}(j),\protect%
\theta _{S}(j),\protect\phi _{S}(j)\right] $ for the vector-density arrows $%
\vec{S_{j}}$. (a3, b3, c3) Distributions of the axis lengths $l_{T}^{n}\left(
j\right) $ ($n=a,b,c$) of the spin-tensor density ellipsoids $T_{j}$.
(a4, b4, c4) Distributions of the orientational Euler angles $\protect\theta %
_{T}$, $\protect\phi _{T}$, $\protect\phi _{T}^{\prime }$ of the spin-tensor
density ellipsoids $T_{j}$. We set $\Omega _{31}/t=0$ for (a1)-(a4), $\Omega
_{31}/t=12.3$ for (b1)-(b4), and $\Omega _{31}/t=19$ for(c1)-(c4). In all
subfigures, we have $\Omega _{12}/t=12.3$, $U/t=0$, $L=32$, and $n=1$. }
\label{fig:magnetism}
\end{figure*}

\section{Order parameters}

\label{order parameters}

The many-body topological properties can be well described
by the degeneracy in entanglement spectrum, entanglement entropy, chemical
potential spectrum, and excited energy gap. The entanglement spectrum is
defined as \cite{Li2008}%
\begin{equation}
\xi _{i}=-\ln (\rho _{i}),
\end{equation}%
with $\rho _{i}$ being the eigenvalue of the reduced density matrix $\hat{%
\rho}_{A}=\mathrm{Tr_{B}|\psi \rangle \langle \psi |}$, where $|\psi \rangle
$ is the ground-state wavefunction and $A,B$ correspond to the left or the
right half of the one-dimensional chain. The system is topological if the
entanglement spectrum is degenerate since the entanglement spectrum
resembles the energy spectrum of edge excitations and vice versa \cite%
{Zhao2015,Yoshida2014,Turner2011,Pollmann2010,
Fidkowski2010,Flammia2009,Li2008}. The quantum criticality of the
interaction-driven topological phase transition can be governed by the von
Neumann entropy \cite%
{Flammia2009,Hastings2010,Daley2012,Abanin2012,jiang2012,Islam2015}
\begin{equation}
S_{\mathrm{vN}}=-\mathrm{Tr_{A}[\hat{\rho}_{A}\log \hat{\rho}_{A}]}.
\end{equation}%
The divergence of the von Neumann entropy at the critical point not only
indicates a continuous transition but also yields a central charge, which
reflects the universality class of phase transition. The von Neumann entropy
of a subchain of length $l$ is given by
\begin{equation}
S_{\mathrm{vN}}=\frac{C}{6}\ln \left[ \sin \frac{\pi l}{L}\right] +\text{%
const},
\end{equation}%
in which the slope at large distance gives the central charge $C$ of the
conformal field theory underlying the critical behavior~\cite%
{centralcharge1,centralcharge2}.

The appearance of edge states is usually considered to be a hallmark of
topological properties for the bulk system. The topological insulator of the
synthetic Hall tube has two gapless edge states inside the gap between the
lowest and the upper branches in the chemical potential spectrum~\cite%
{Barbarino2018}, which is essentially the energy required to add an atom to
a system of $N$\ atoms and can be defined as
\begin{equation}
\mu =E_{g}^{o}(N)-E_{g}^{o}(N-1).
\end{equation}%
Here, $E_{g}^{o}(N)$\ is the ground-state energy of $N$\ atoms under open
boundary condition. The topological ground state of the synthetic Hall tube
is nondegenerate and separated from the first excited state by a finite gap,
which closes and reopens in the process of topological phase transition \cite%
{Han2019}. The excited energy gap is defined as%
\begin{equation}
\Delta _{e}=E_{e}^{p}(N)-E_{g}^{p}(N),
\end{equation}%
where $E_{e}^{p}(N)$ [$E_{g}^{p}(N)$] is the first-excited (ground) state
energy of $N$ atoms under periodic boundary condition.

Due to the coexistence of the spin-vector and -tensor Zeeman fields, the
magnetism of the synthetic Hall tube should be described by whole spin-1
local magnetic orders (8 spin-moments with 3 spin-vectors and 5
spin-tensors)~and their correlations \cite{zhou2020}. The local spin-vector
\begin{equation}
\vec{S}_{j}=(\langle S_{j}^{x}\rangle ,\langle S_{j}^{y}\rangle ,\langle
S_{j}^{z}\rangle )^{T},
\end{equation}%
while the local spin-tensor fluctuation matrix $T_{j}\ $has tensor moments
\begin{equation}
T_{j}^{\alpha \beta }=\langle \{S_{j}^{\alpha },S_{j}^{\beta }\}\rangle
/2-\langle S_{j}^{\alpha }\rangle \langle S_{j}^{\beta }\rangle .  \label{TE}
\end{equation}%
Geometrically, $\vec{S}_{j}$ is characterized by an arrow and $T_{j}$ is governed by an
ellipsoid (with principle axis lengths $l_{T}^{n}\left( j\right) $ ($n=a,b,c$%
) and orientations $\vec{v}_{T}^{n}\left( j\right) $ given by the
square-roots of the eigenvalues and eigenvectors of $T_{j}^{\alpha \beta }$~%
\cite{Bharath}). Since the magnetic properties are mainly determined by $%
\hat{H}_{\Omega }$, all the insulators are spiral paramagnetic phases
without any long-range correlations. As a result, eight independent
geometric parameters, including the length $l_{S}$ and spherical coordinates $%
\theta _{S}$, $\phi _{S}$ of the arrow, the two axis lengths $l_{T}^{a,b}$
with the third axis length $l_{T}^{c}=\sqrt{{2-(l_{S})}^{2}-{(l_{T}^{a})}%
^{2}-{(l_{T}^{b})}^{2}}$, and the orientational Euler angles $\theta _{T}$, $%
\phi _{T}$, $\phi _{T}^{\prime }$ of the ellipsoid, are chosen to
quantitatively characterize and geometrically visualize the magnetic orders.\newline

\section{Quantum phases}

\label{Many-body phases}

\subsection{Noninteracting case ($U=0)$}

\label{U=0}

We first address the case of the noninteracting case ($U=0$). For $\Omega
_{-}<\Omega _{31}<\Omega _{+}$, the ground state is a topological insulator
and vice versa. As a result, we can discuss the magnetisms of the
topological and nontopological insulators as varying the spin-tensor Zeeman
field $\Omega _{31}$ for a fixed spin-vector Zeeman field $\Omega
_{12}/t=12.3$. When $\Omega _{31}=0$, the system is the same as the spin-1
spin-orbit coupled optical lattice only with the spin-vector Zeeman field
\cite{Natu2015,Barbarino2015,spin-1SOC2016,SOC-1BEC1}. In this case, the
spin-vector arrow has a unit length $l_{S}=1$ and spirals in the $x-y$ plane
(i.e., $\theta _{S}$ is a constant and $\phi _{S}$ changes cyclically), as
shown in Figs.~\ref{fig:magnetism}(a1) and \ref{fig:magnetism}(a2). The
spin-tensor ellipsoid almost is a plate with large $l_{T}^{b,c}$ and small $%
l_{T}^{a}$ [see Fig.~\ref{fig:magnetism}(a3)], and also spirals with a
cyclical variation $\phi _{T}$ and constants $\theta _{T}$, $\phi
_{T}^{\prime }$ [see Fig.~\ref{fig:magnetism}(a4)], since the spin-tensor
ellipsoid depends crucially on the three spin-vector operators $S^{\alpha }$
[see Eq.~(\ref{TE})]. This paramagnetic insulator dominated only by the
spin-vector is called nontopological spin-vector paramagnetic insulator
(NTSV). For $\Omega _{31}<\Omega _{-}$ (i.e., a small spin-tensor Zeeman
field), the ground state is still the NTSV.

For the topological regime with $\Omega _{-}<\Omega _{31}<\Omega _{+}$, the
spin-vector arrows also spiral in the $x-y$ plane but have short lengths $%
l_{S}<1$, as shown in Figs.~\ref{fig:magnetism}(b1) and \ref{fig:magnetism}%
(b2). In this case, the spin-vector cannot fully describe the magnetic
properties and the spin-tensor should be considered. The spin-tensor
ellipsoids have finite $l_{T}^{a,b,c}$ [see Fig.~\ref{fig:magnetism}(b3)],
and also spiral in the $x-y$ plane with a cyclical variation $\phi _{T}$ and
constants $\theta _{T}$, $\phi _{T}^{\prime }$ [see Fig.~\ref{fig:magnetism}%
(b4)]. Different from the NTSV, this spiral spin-tensor ellipsoids only
depend on the spin-tensor Zeeman field. This topological insulator is called
topological spin-mixed paramagnetic insulator (TSM). For $\Omega
_{31}>\Omega _{+}$ (i.e., a large spin-tensor Zeeman field), the spin-vector
arrow vanishes (i.e., $l_{S}=0$), as shown in Figs.~\ref{fig:magnetism}(c1)
and \ref{fig:magnetism}(c2). In this case, the magnetic orders are fully
dominated by the spin-tensor ellipsoid. The ellipsoids have $l_{T}^{b,c}\sim
1$ and $l_{T}^{a}\rightarrow 0$ [see Fig.~\ref{fig:magnetism}(c3)], and also
spiral in the $x-y$ plane with a cyclical variation $\phi _{T}$ and
constants $\theta _{T}$, $\phi _{T}^{\prime }$ [see Fig.~\ref{fig:magnetism}%
(c4)]. This paramagnetic insulator without the spin-vector is called
nontopological spin-tensor paramagnetic insulator (NTST).

\begin{figure}[t]
\centering
\includegraphics[width = 3.5in]{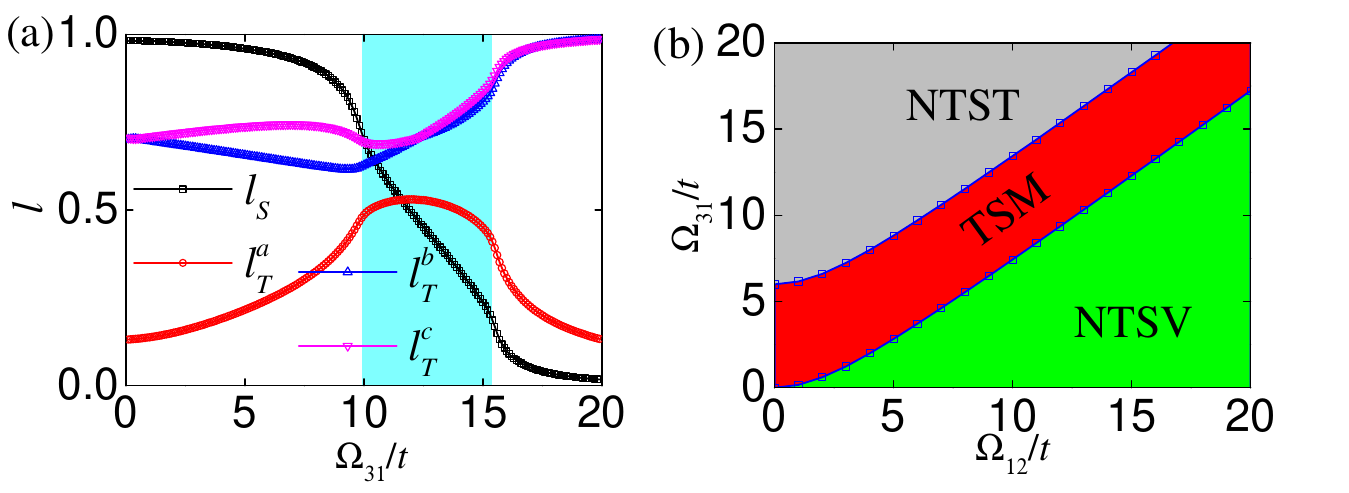} %
\hskip 0.0cm
\caption{(a) The length of the spin-vector arrow $l_{S}$ and the axis
lengths of the spin-tensor ellipsoid $l_{T}^{a,b,c}$ as functions of the
spin-tensor Zeeman field $\Omega _{31}/t$ with a spin-vector Zeeman field $%
\Omega _{12}/t=12.3$. (b) Phase diagram in the $\Omega _{12}-\Omega _{31}$
plane. In all subfigure, we have $U=0$, $L=32$, and $n=1$.}
\label{fig:phasediagnoninter}
\end{figure}

The above analysis of Fig.~\ref{fig:magnetism} shows that there exist two
topologically magnetic phase transitions as increasing the spin-tensor Zeeman field~$\Omega _{31}/t$. One
is the transition from the NTSV to the TSM at $\Omega _{31}^{c1}=\Omega _{-}$%
. At this critical point, the spin-vector arrow length $l_{S}$ drops
rapidly, but the ellipsoid's axis lengths $l_{T}^{a,b,c}$ increase rapidly
[see Fig.~\ref{fig:phasediagnoninter}(a)]. The other is the transition from
the TSM to the NTST at $\Omega _{31}^{c2}=\Omega _{+}$. At this critical
point, the spin-vector arrow length $l_{S}$ suddenly becomes zero, and the
ellipsoid's axis lengths $l_{T}^{a,b}$ ($l_{T}^{c}$) increase (decrease)
abruptly [see Fig.~\ref{fig:phasediagnoninter}(a)]. Figure~\ref%
{fig:phasediagnoninter}(b) shows the phase diagram in the $\Omega
_{31}-\Omega _{12}$ plane. Both the phase transitions of NTSV$%
\leftrightarrow $TSM and TSM$\leftrightarrow $NTST [see blue lines in Fig.~%
\ref{fig:phasediagnoninter}(b)] are of second-order with a closing excited
energy gap $\Delta _{e}$ at the critical points \cite{Barbarino2018,Han2019}.

\subsection{Interacting case ($U>0$)}

\label{U>0}

\begin{figure}[t]
\centering
\includegraphics[width = 3.5in]{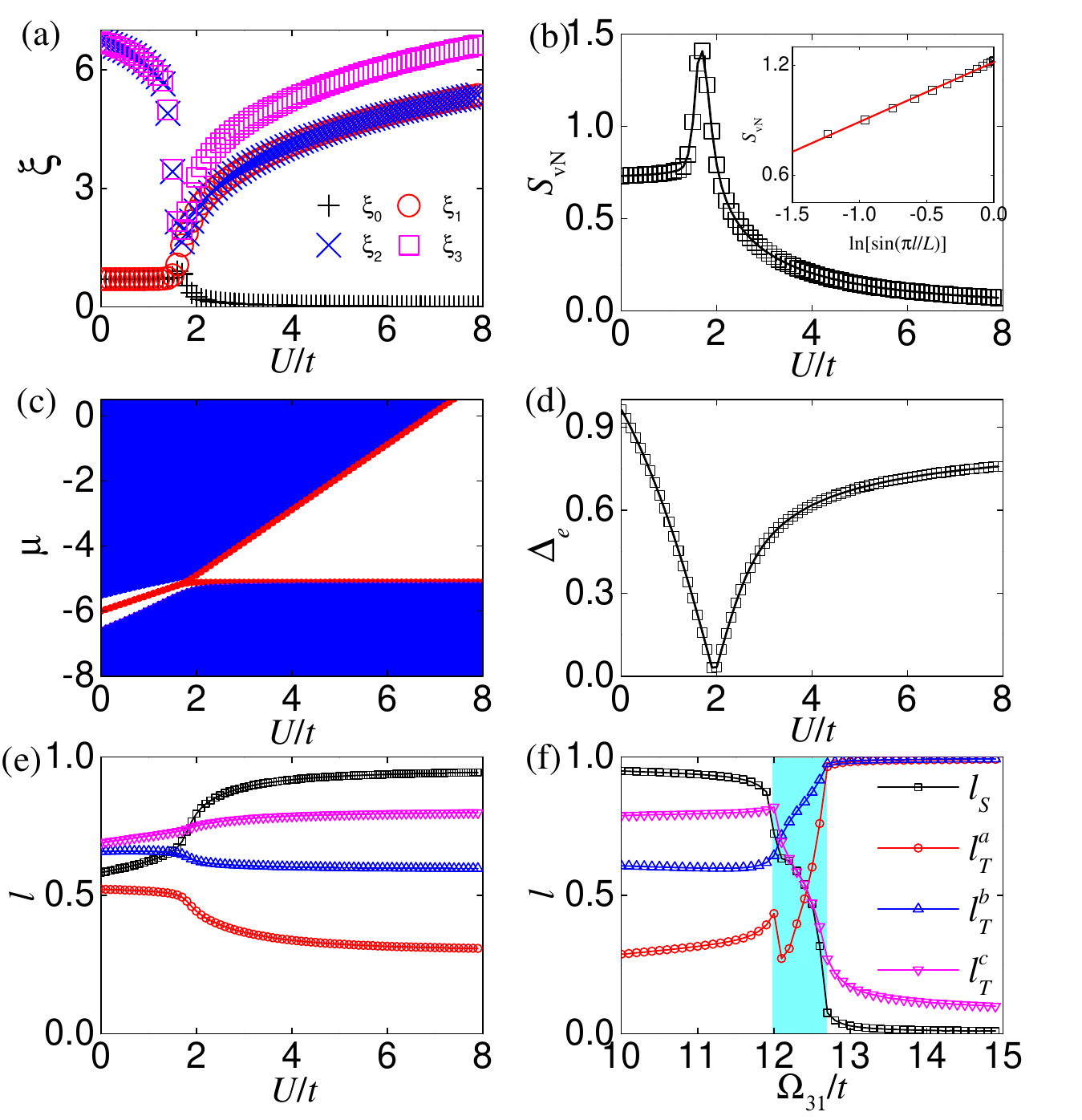} \hskip 0.0cm
\caption{(a) The lowest four levels in the entanglement spectrum $\protect%
\xi _{i}$ ($i=0,1,2,3$), (b) the von Neumann entropy $S_{\mathrm{vN}}$, (c)
the chemical potential spectrum $\protect\mu $, (d) the excited energy gap $%
\Delta _{e}$, and (e) the length of the spin-vector arrow $l_{S}$ and the
axis lengths of the spin-tensor ellipsoid $l_{T}^{a,b,c}$ as functions of
the interaction strength $U/t$. In (a, b, c, e), $\Omega _{31}/t=11$ and $L=32$%
, and open boundary condition is used. In (d), $\Omega _{31}/t=11$ and $L=12$%
, and periodic boundary condition is used. The inset of (b) shows that the
von Neumann entropy of a subchain of length $l$ as a function of $\sin (%
\protect\pi l/N)$ for a chain with $L=32$ at the critical point $%
U_{c}/t=1.85$. The solid line is the linear fit: $S_{\mathrm{vN}}={\frac{C}{6%
}}\ln [\sin (\protect\pi l/N)]+1.22$ with $C\approx 2$. The central charge
is six times the slope of the linear fit. (f) The length of the spin-vector
arrow $l_{S}$ and the axis lengths of the spin-tensor ellipsoid $%
l_{T}^{a,b,c}$ as functions of the spin-tensor Zeeman field $\Omega _{31}/t$
with the interaction strength $U/t=6$. In all subfigures, we have $\Omega _{12}/t=12.3$ and $n=1$.}
\label{fig:entanglement}
\end{figure}

We now explore many-body properties induced by the repulsive
interaction ($U>0$). We first address the topological properties driven by
interaction, when $\Omega _{12}/t=12.3$ and $\Omega _{31}/t=11$. For a weak
interaction, the entanglement spectrum $\xi _{i}$ is two-fold degeneracy,
and no longer degenerate beyond a critical interaction strength $U_{c}/t\sim
1.85$, as shown in Fig.~\ref{fig:entanglement}(a). Without any symmetry
breaking in this processing, it is a typical topological phase transition
from a topological insulator to a nontopological insulator. As demonstrated
in Fig.~\ref{fig:entanglement}(b), sharp features of the von Neumann
entropy $S_{\mathrm{vN}}$ emerge at the critical point. From inset of Fig.~\ref{fig:entanglement}(b),
we estimate $C\sim 1.97$, which is close to the universality class of the
Luttinger liquid ($C=2$) and shows the continuous of the topological phase
transition. On the other hand,
in the absence of interaction, there are two gapless edge states inside the
bulk band gap in the chemical potential spectrum $\mu $. As increasing the
interaction strength beyond a critical value $U_{c}/t$, these edge states
merge into the bulk band and the system becomes a nontopological insulator,
as shown in Fig.~\ref{fig:entanglement}(c). Moreover, the excited energy gap
$\Delta _{e}$ closes at the same critical value and then reopens, as shown
in Fig.~\ref{fig:entanglement}(d). This critical point is well consistent
with that derived from entanglements in Figs.~\ref{fig:entanglement}(a) and %
\ref{fig:entanglement}(b).

We now explore the magnetic orders in the presence of interaction. Based on
the above graphics of Fig.~\ref{fig:magnetism}, it can be found that the
spin-vector arrows and the spin-tensor ellipsoids exhibit the same spiral
features, but show the distinct lengths $l_{S}$ and $l_{T}^{a,b,c}$. This
means that these lengths are adequate to describe the magnetic properties.
As a result, we only calculate the lengths $l_{S}$ and $l_{T}^{a,b,c}$ and
ignore the angles $\theta _{S}$, $\phi _{S}$, $\theta _{T}$, $\phi _{T}$,
and $\phi _{T}^{\prime }$ hereafter.

In Fig.~\ref{fig:entanglement}(e), we
plot the lengths $l_{S}$ and $l_{T}^{a,b,c}$ as functions of the interaction
strength $U/t$. This figure shows that at the critical point $U_{c}/t$, the
spin-vector arrow length $l_{S}$ suddenly increases to $l_{S}\sim 1$, and the
spin-tensor ellipsoid's axis lengths $l_{T}^{a,b}$ drop rapidly, which
indicate that this topologically magnetic phase transition from the TSM to
the NTSV occurs. In Fig.~\ref{fig:entanglement}(f), we plot the lengths $%
l_{S}$ and $l_{T}^{a,b,c}$ as functions of the spin-tensor Zeeman field
strength $\Omega _{31}$ for a large interaction strength $U/t=6$. In this
case, all the insulators are nontopological since the entanglement spectrum $%
\xi _{i}$ is nondegenerate. Interestingly, as increasing the spin-tensor
Zeeman field $\Omega _{31}$, the vector length $l_{S}$ and the ellipsoid's
axis lengths $l_{T}^{a,b}$ firstly remain, then $l_{S}$ rapidly decreases to
$l_{S}<1$ and $l_{T}^{a,b}$ ($l_{T}^{c}$) increase (decrease) rapidly. The
corresponding phase is called the nontopological spin-mixed paramagnetic
insulator (NTSM). Further increasing the spin-tensor Zeeman field $\Omega _{31}$, $l_{S}$ suddenly drops
to $l_{S}\sim 0$, and $l_{T}^{a,b}$ rapidly increase to $l_{T}^{a,b}\sim 1$,
i.e., the system enters into the NTST. Note that for a small interaction,
the fundamental properties are similar to that in Fig.~\ref%
{fig:phasediagnoninter}(a) and thus not plotted here.

\begin{figure}[t]
\includegraphics[width = 3.6in]{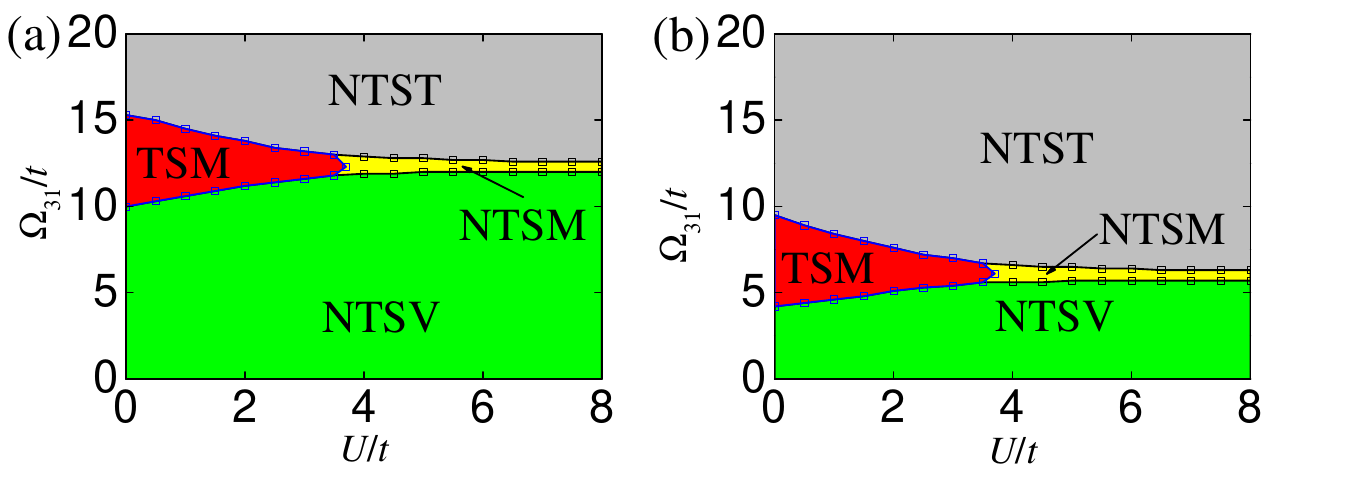} \hskip 0.0cm
\centering
\caption{Phase diagrams in the $\Omega _{31}-U$ plane for the different
spin-vector Zeeman fields $\Omega _{12}/t=12.3$ (a) and $\Omega _{12}/t=6.0$
(b). The blue and black lines show the continuous phase transitions. The
blue lines denote the liquids with $C=2$. In all subfigure, we have $L=32$
and $n=1$.}
\label{fig:phasediag}
\end{figure}

Finally, with the help of the calculated entanglement spectrum, entanglement
entropies, energy gaps, and local magnetic orders, in Fig.~\ref%
{fig:phasediag} we map out phase diagrams the $\Omega _{31}-U$ plane for the
different spin-vector Zeeman fields $\Omega _{12}/t=12.3$ (a) and $\Omega
_{12}/t=6.0$ (b). This figure shows clearly four different phases such as
the TSM, the NTSM, the NTSV, and the NTST, which are well controlled by both
the spin-vector and -tensor Zeeman fields as well as the repulsive interaction. Moreover, all the phase
transitions with a closing excited energy gap $\Delta _{e}$ are of second
order. \newline


\section{Conclusions}
\label{Summary}

Before ending up this paper, we briefly discuss how to observe these quantum
phases and phase transitions in cold atom experiments. The entanglement entropy can be measured using
quantum interference of many-body twins of ultracold atoms in optical lattices~\cite%
{Islam2015}. The excited energy gap closing in the processing of topological phase
transition can be observed via momentum-resolved analysis of the quench
dynamics~\cite{Han2019}. The local magnetic orders can
be measured by isolating the sites of interest using additional
site-resolved potentials~\cite{Greiner2011,Greiner2016,Greiner2017,Bloch2016}. Thus, all the quantum phases and phase transitions can be observed in current experimental setups.

In conclusion, we have studied the many-body physics of interacting synthetic Hall tube by the state-of-the-art DMRG numerical method. We have found four quantum phases, including the TSM, the NTSM, the NTSV, and the NTST, by means of the calculated entanglement spectrum, entanglement entropies, energy gaps, and local magnetic orders. These quantum phases depend crucially on the interaction and the spiral spin-vector and -tensor Zeeman fields induced by the inter-leg couplings. Our work paves a new way to explore many-body (topological) states induced by both the spiral spin-vector and -tensor Zeeman fields.\newline

\section*{Acknowledgments}

This work is supported by the National Key R\&D Program of China under Grant
No.~2017YFA0304203; the NSFC under Grant No.~11674200;
the Fund for Shanxi `1331 Project' Key Subjects Construction; and Research Project Supported by Shanxi Scholarship Council of
China.

%

\end{document}